\documentclass{ifacconf}

\usepackage{amsmath}
\usepackage{algorithmic}
\usepackage{array}
\usepackage{color, soul} 
\usepackage{amssymb}
\usepackage{graphicx}
\usepackage{subfig}
\usepackage{natbib} 
\usepackage{eurosym} 
\usepackage[width=\textwidth]{caption} 

\begin{document}
\begin{frontmatter}

\title{Conflict-free Charging and Real-time Control for an Electric Bus Network 
\thanksref{footnoteinfo}} 

\thanks[footnoteinfo]{The paper has been supported by the Swedish Energy Agency through the project “Operational Network Energy Management for Electrified Buses” (46365-1) and IRIS with Transport Area of Advance (Chalmers University of Technology).}

\author[First]{Rémi Lacombe} 
\author[First]{Nikolce Murgovski}
\author[Second]{Sébastien Gros} 
\author[First]{Balázs Kulcsár}

\address[First]{Department of Electrical Engineering, Chalmers University of Technology, Gothenburg, Sweden \\ (e-mail: $\{$lacombe,nikolce.murgovski,kulcsar$\}$@chalmers.se)}
\address[Second]{Department of Engineering Cybernetics, Norwegian University of Science and Technology, Trondheim, Norway \\ (e-mail: sebastien.gros@ntnu.no)}

\begin{abstract} 
The rapid adoption of electric buses by transit agencies around the world is leading to new challenges in the planning and operation of bus networks. In particular, the limited driving range of electric vehicles imposes operational constraints such as the need to charge buses during service. Research on this topic has mostly focused on the strategic and tactical planning aspects until now, and very little work has been done on the real-time operational aspect. To remedy this, we propose integrating the charging scheduling problem with a real-time speed control strategy in this paper. The control problem is formulated as a mixed-integer linear program and solved to optimality with the branch-and-bound method. Simulations are carried out by repeatedly solving the control problem in a receding horizon fashion over a full day of operation. The results show that the proposed controller manages to anticipate and avoid potential conflicts where the charging demand exceeds the charger capacity. This feature makes the controller achieve lower operational costs, both in terms of service regularity and energy consumption, compared to a standard first-come, first-served charging scheme.
\end{abstract}


\end{frontmatter}

\section{Introduction}

The past few years have seen several municipalities make pledges to transition to fully electric bus fleets over the current decade (\cite{bloomberg}). Owing to having no tailpipe emissions and to their lower energy consumption (\cite{lajunen}), electric buses are being deployed at a fast pace in urban centers. This rapid evolution comes with its fair share of technical challenges, however, as transit agencies are faced with new problems in the bus planning process. These challenges occur at the strategic planning stage, with the fleet and charging infrastructure investment problem (\cite{pelletier}) and the charging infrastructure placement problem (\cite{xylia}), as well as at the tactical planning stage, with the electric vehicle scheduling problem (\cite{tang}) and the charging scheduling problem (\cite{he}).

Most papers in the literature dealing with the tactical planning stage focus on the electric vehicle scheduling problem, i.e. the assignment of vehicles to trips planed according to a static timetable. As emphasized in the recent review paper \cite{perumal}, very few works in this line of research have addressed the question of robust scheduling of electric buses, and fewer still the question of their real-time control at the operational level. This is however a crucial aspect since urban environments are typically highly uncertain and external disturbances can severely disturb bus service. Electric buses are particularly vulnerable to this due to their limited driving range and frequent charging needs. 

Using the terminology found in the public transit literature (\cite{ibarra}, \cite{perumal}), we propose in this paper an integrated approach including the charging scheduling and real-time operational control problems of the electric bus planning process. Here, charging scheduling is to be understood in a general sense as the problem of deciding when to charge an electric vehicle. To the best of the authors' knowledge, no other work has yet attempted to integrate a real-time control strategy with the specific charging constraints faced by electric buses in operation.

Following the main trend in electric bus planning research (\cite{perumal}), the charging technology considered in this paper is that of fast pantograph chargers. These chargers deliver a high charging power and can either be installed at some of the bus stops, or at a neighboring bus depot. In case of a limited number of chargers, the charging demand from the vehicles may be greater than the installed charging capacity, resulting in charging \textit{conflicts} that can be damaging to bus operations. Real-time bus control is carried out in this work through a speed control strategy combined with bus holding at the chargers location (\cite{ibarra}). The control problem, where control decisions are taken over an horizon, is formulated in the mathematical programming framework as a mixed-integer linear program (MILP). 

The main contribution of this paper is to be the first work where the charging scheduling problem for electric buses is integrated with a real-time control strategy. This means that the main operational dynamics of a bus line, such as higher passenger accumulations at stops for delayed buses, are now integrated to the charging decisions. This aspect in particular has been shown in simulation studies (\cite{hans}) to be the single most important operational feature in a bus line model and should therefore not be overlooked when considering robust bus scheduling. In addition, partial charging is allowed and charging times can take any real values instead of being constrained to a discrete set (e.g. only multiples of a given time interval), contrary to most works in the charging scheduling literature.

The article is structured as follows. Section 2 presents a bus network model and the main modeling assumptions under which it is constructed. This model is then used in Section 3 to assemble the charging scheduling and real-time control problem as an MILP. Simulation results from a case study with two bus lines and based on real data are presented in Section 4. Finally, the article ends with some concluding remarks in Section 5.

\section{Modeling}
 
This section introduces a model including both the operational and charging aspects of an electric bus network.
 
\subsection{General assumptions}

It is assumed that a prior tactical planning phase has been carried out to fix the number of vehicles in operation and the desired service headway for each bus line. Note that a timetable-based bus system could also be considered at the price of minor modifications only, but we choose to focus on headway regularity in this paper. In addition, it is assumed that all fast chargers are located at a single terminal common to all bus lines and are shared by all buses. This shared terminal is the only stop that any two bus lines have in common, and is also the only stop at which bus holding can be applied.

The following assumptions are made on bus operations:
\begin{itemize}
    \item Each vehicle can only be dispatched on a single bus line: buses are not allowed to be redeployed during service.
    \item The bus order in each line is fixed: no overtaking is allowed between buses of the same line.
    \item Bus capacity constraints are not considered.
\end{itemize}
These assumptions are necessary to keep a tractable problem in the present work, but note that they are relatively common in the literature. See \cite{gkiotsalitis3} and references therein.


\begin{figure}
\begin{center}
\includegraphics[width=7.5cm]{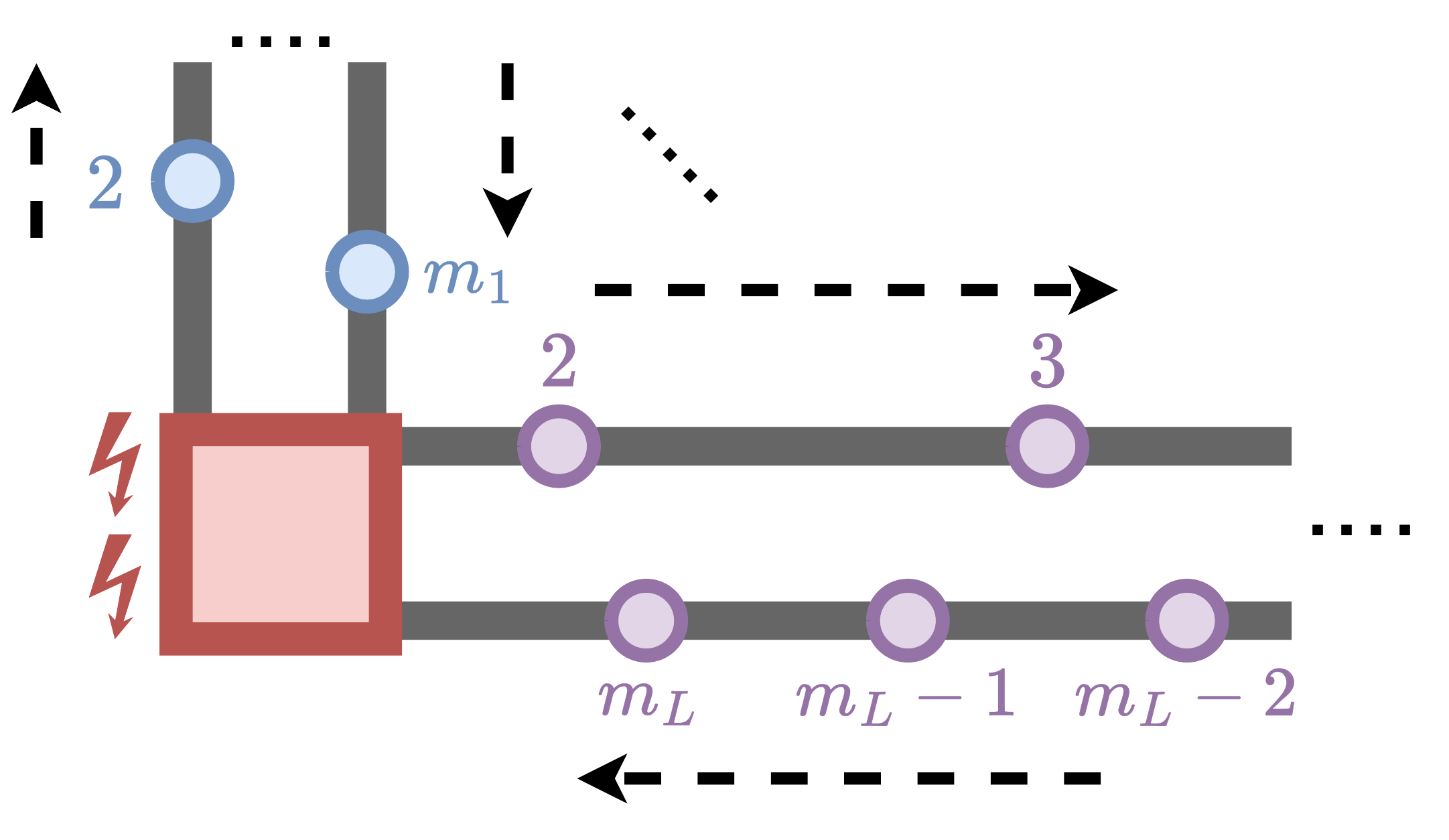} 
\caption{Representation of the bus network structure considered. Only two bus lines (with indices $1$ and $L$) are represented explicitly. The terminal, where all chargers are located, is represented in red. A few regular bus stops are also displayed as circles. Black arrows indicate the direction of driving.}
\label{fig:network}
\end{center}
\end{figure}

Let us consider a bus network with $L$ separate lines and let ${\mathcal{L}=\{1,...,L\}}$ be the set of line indices. For any bus line ${l\in\mathcal{L}}$, let ${\mathcal{I}_l=\{1,...,n_l\}}$ and ${\mathcal{J}_l=\{1,...,m_l\}}$ be the sets of bus and stop indices, respectively. The stop with index 1 is the shared bus terminal. Note that the structure of each bus line can be either circular or linear with this choice of notation. For the latter case, each stop (except the terminal) is simply assigned two distinct indices. An example of the bus network structure considered in this paper is given in Fig. \ref{fig:network}.

\subsection{Bus dynamics}

The following expressions hold for any ${l\in\mathcal{L}}$, ${i\in\mathcal{I}_l}$ and ${j\in\mathcal{J}_l}$ for the rest of this section, unless specified otherwise. Without loss of generality, the bus with index $n_l$ for line $l$ is also referred to as the bus with index 0.

We present here a simplified operational bus line model where the speed control strategy is implemented by choosing a single speed command on each inter-stop link. This average bus speed can be equivalently expressed as a travel time command since inter-stop distances are known. Let then $\tau_{l,i,j}$ be a \textit{control input} denoting the desired travel time of bus $i$ from line $l$ on the $j$-th link of line $l$. It is assumed that:
\begin{equation} \label{eq:time_bounds}
    T^{\text{min}}_{l,j} \leq \tau_{l,i,j} \leq T^{\text{max}}_{l,j},
\end{equation}
where $T^{\text{min}}_{l,j}$ and $T^{\text{max}}_{l,j}$ are the minimum and maximum travel times, respectively, that are allowed on that link. These bounds may for example reflect the legal speed limit, or a minimum acceptable speed below which buses would be disturbing the surrounding traffic too much. 

Let ${t_{l,i,j}}$ be a \textit{state} variable associated with the bus of index $i$ from line $l$ and denoting the time instant at which this bus arrives at the stop with index $j$. The no-overtaking assumption is then expressed as:
\begin{equation} \label{eq:no_overtaking}
    t_{l,i,j}-t_{l,i-1,j} \geq 0,
\end{equation}
meaning that the preceding bus must always arrive at the stop first.


Next, the evolution of this state variable can be modeled as:   
\begin{equation} \label{eq:time_dynamics}
    t_{l,i,j+1} = t_{l,i,j} + \tau_{l,i,j} + d_{\text{pax}}\lambda_{l,j}(t_{l,i,j}-t_{l,i-1,j}), \quad j>1, 
\end{equation}
where $j+1$ corresponds to the stop with index 1 for the case $j=m_l$. The case $j=1$ is treated later since it involves a visit to the terminal. The last term in the right-hand side of this equation is the dwell time at stop $j$, where $d_{\text{pax}}$ is the mean passenger boarding time and $\lambda_{l,j}$ is the mean passenger arrival rate at the $j$-th stop of line $l$. The dwell time is assumed to be equal to the time needed for the passenger exchange since no holding is allowed at intermediary stops. We make the standard assumption that the duration of the passenger exchange is dominated by the boarding operation (\cite{hans}). Therefore, the dwell time is equal to the boarding time of the passengers that have accumulated at stop $j$ since the last visit of the preceding bus (with index $i-1$). 

The case of the terminal is a bit special since it is assumed that bus holding can be carried out there (when waiting for a fast charger to become available for example). The holding time of bus $i$ at the terminal is modeled by the control input $w_{l,i}$. It is assumed that:
\begin{equation} \label{eq:holding_bound}
    w_{l,i} \geq d_{\text{pax}}\lambda_{l,1}(t_{l,i,1}-t_{l,i-1,1}),
\end{equation}
in order to allocate enough holding time for the passenger exchange to take place. Other than that, buses are allowed to wait at the terminal for as long as they want. 

\subsection{Charging modeling}

Let $\sigma_{l,i,j}$ be a state variable for the state-of-charge of bus $i$ from line $l$ when arriving at the stop with index $j$. Its evolution is governed by:
\begin{equation} \label{eq:SoC_dynamics}
    \sigma_{l,i,j+1} = \sigma_{l,i,j} - \frac{E_{l,i,j}}{Q},\quad j > 1, 
\end{equation}
where again $j+1$ denotes the stop with index 1 when ${j=m_l}$. In this equation, $E_{l,i,j}$ is the energy consumption of bus $i$ on the $j$-th link of line $l$, and $Q$ is the total battery capacity of the bus. The buses could have different battery capacities in principle, but we present the case where all vehicles have the same capacity $Q$ here for the sake of simplicity (e.g. if all electric buses are of the same model). 

In this paper, the energy consumption is modeled as a linear function of the travel time, i.e. $E_{l,i,j}$ is a linear function of $\tau_{l,i,j}$. This is done by using the detailed electric motor and battery models from \cite{lacombe} and fitting a linear function to the values obtained. Note, however, that \eqref{eq:SoC_dynamics} is a general formulation which is suitable for other energy consumption models as well. For instance, one could also consider a simpler model with a constant energy consumption $E_{l,i,j}$ on each link.

In order to model charging at the terminal, additional control inputs are introduced. Let $b_{l,i}$ be a binary variable taking value 1 if bus $i$ from line $l$ decides to charge at the terminal, and 0 otherwise. Similarly, let the continuous variable $c_{l,i}$ denote the time spent charging. Then, using the big-M method, the charging time is constrained as: 
\begin{equation}
    0 \leq c_{l,i} \leq M b_{l,i},
\end{equation}
where $M$ is a large real number. Assuming linear charging profiles, the evolution of the state-of-charge at the terminal and upon reaching the next stop is:
\begin{equation}
    \sigma_{l,i,2} = \sigma_{l,i,1} + \frac{P_{\text{char}}}{Q}c_{l,i} - \frac{E_{l,i,1}}{Q},
\end{equation}
where $P_{\text{char}}$ is the power delivered by the fast charger. Note that 
taking a decision ${b_{l,i}=0}$ thus results in an unchanged state-of-charge when leaving the terminal.

In addition, the state-of-charge of every bus must be larger than a certain threshold value $\sigma^{\text{min}}_l$ when leaving the terminal in order to guarantee the feasibility of the next trip:
\begin{equation} \label{eq:SoC_feasibility}
    \sigma^{\text{min}}_l \leq \sigma_{l,i,1} + \frac{P_{\text{char}}}{Q}c_{l,i} \leq 1. 
\end{equation}
This threshold could also be set so as to limit the depth of discharge allowed for the bus batteries and thus improve their cycle life.

Finally, the time update at the terminal can be modeled as:
\begin{equation} \label{eq:time_last}
    t_{l,i,2} = t_{l,i,1} + w_{l,i} + c_{l,i} + 2d_{\text{char}}b_{l,i} + \tau_{l,i,1},
\end{equation}
where $d_{\text{char}}$ is the time needed to begin or end the charging operation. Note that this formulation can cover several network structures beside the one considered in this paper. Large values of $d_{\text{char}}$ could for example model a scenario where the fast chargers are not directly located at the line terminal, e.g. if they are at a nearby bus depot.

\subsection{Charger exclusion constraints}

Inter-line coupling constraints are now considered. With the modeling assumptions presented above, the only source of coupling between bus lines comes from sharing the same chargers at the terminal. Additional constraints must be introduced to ensure that no two buses are ever using the same charger simultaneously.

Here, we present the case where a single fast charger is located at the terminal, for the sake of clarity. However, note that the case with several chargers would be identical with only extra decision variables needed to model the additional chargers.

Let ${t^{\text{char}}_{l,i}}$ be the time at which bus $i$ from line $l$ starts using the charger (in case of a positive charging decision). With the notations introduced previously:
\begin{equation}
    t^{\text{char}}_{l,i} = t_{l,i,1} + w_{l,i} + d_{\text{char}}. 
\end{equation}
The bus then stops using the charger at time ${t^{\text{char}}_{l,i}}+c_{l,i}$.

For all pairs of buses ${\{l,i\}\in\mathcal{L}\times\mathcal{I}_l}$ and ${\{l',i'\}\in\mathcal{L}\times\mathcal{I}_{l'}}$, non-simultaneous charger use can be encoded in the following Boolean expression:
\begin{subequations} \label{eq:boolean}
\begin{align} 
(t^{\text{char}}_{l,i} + c_{l,i} \leq t^{\text{char}}_{l',i'}) \oplus (&t^{\text{char}}_{l',i'} + c_{l',i'} \leq t^{\text{char}}_{l,i}) \nonumber \\
& \lor (1-b_{l,i}) \lor (1-b_{l',i'}), \tag{\ref{eq:boolean}}
\end{align}
\end{subequations}
where $\oplus$ stands for the classical \texttt{xor}, or exclusive \texttt{or}, logical operation. In short, a problem can only occur if both buses have taken positive charging decisions. If such is the case, either of the two buses must finish using the charger before the other starts using it.

This Boolean expression can be translated into a more practical set of constraints by introducing an auxiliary binary variable $\psi_{l,i,l',i'}$ and using the big-M method. This new decision variable encodes the order in which the two buses use the charger: ${\psi_{l,i,l',i'}=1}$ means that bus $i$ from line $l$ would be charged after bus $i'$ from line $l'$, and vice-versa for ${\psi_{l,i,l',i'}=0}$. Note that this variable is only relevant if both buses take positive charging decisions. An equivalent version of expression \eqref{eq:boolean} is then:
\begin{subequations}
\begin{align}
& t^{\text{char}}_{l,i} + c_{l,i} - t^{\text{char}}_{l',i'} \leq (1-b_{l,i})M + (1-b_{l',i'})M \nonumber \\
& \hspace{5.7cm} + \psi_{l,i,l',i'} M, \label{eq:exclusion_1} \\
& t^{\text{char}}_{l',i'} + c_{l',i'} - t^{\text{char}}_{l,i} \leq (1-b_{l,i})M + (1-b_{l',i'})M \nonumber \\
& \hspace{4.8cm} + (1-\psi_{l,i,l',i'})M. \label{eq:exclusion_2} 
\end{align}
\end{subequations}

\section{Problem Formulation} 

The model assembled in the previous section is now used to express the charging scheduling and real-time control problem in the mathematical programming framework. 

\subsection{Horizon design}

Ideally, we would want to optimize passenger service and operational cost over an entire day of operation, or even longer. However, the resulting optimization problem would be too large with the modeling choices described above, which is why we formulate the control problem over a shorter horizon instead. This problem is then solved repeatedly in a receding horizon fashion over the desired control period.

Let us therefore consider the control problem over a finite time horizon of length $T$. Since the bus dynamics have been expressed in the space domain earlier, with time as a state variable, this common time horizon is mapped to a separate spatial horizon for every bus in the network. The spatial horizon of each bus starts at the current position of this bus, and consists of all the upcoming bus stops that this bus is expected to visit over the next $T$ time units. Some of the buses may visit the same stop several times for long horizons. In this case, a positive index $k$ is used to denote the index of each visit. For instance, the time at which bus $i$ from line $l$ arrives at the $j$-th stop for the $k$-th time on its control horizon is noted $t_{l,i,j}^k$. A similar logic is applied to the rest of the notations hereafter.

\begin{figure}[b]
\begin{center}
\includegraphics[width=8.4cm]{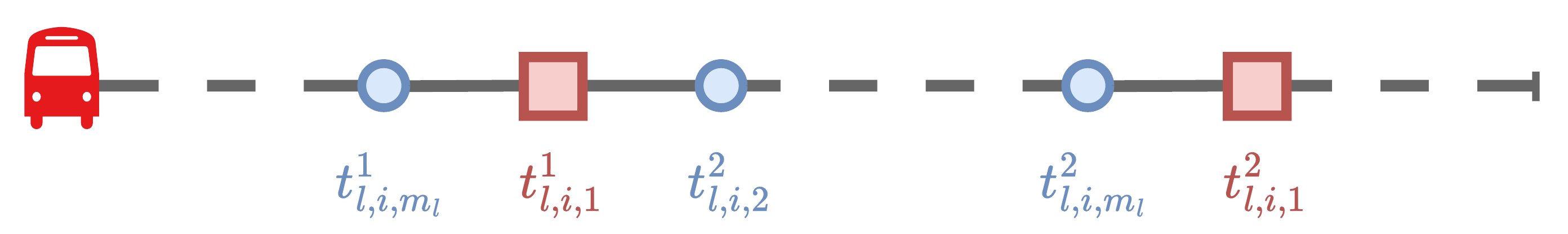}
\caption{Representation of the spatial control horizon for bus $i$ of line $l$. The times at which this bus arrives at each stop are displayed. The red stops represent terminal visits and the blue stops visits to normal stops.} 
\label{fig:horizon}
\end{center}
\end{figure}

An illustration of the control horizon for a generic bus with index $i$ and line index $l$ is given in Fig. \ref{fig:horizon}. The bus is expected to visit the terminal twice on its horizon in this particular example. In general, we recommend choosing $T$ such that several terminal visits are included on the horizon of each bus, so that potential charging conflicts can be anticipated well in advance.

\subsection{MILP}

In order to model the costs of the transit agency associated with the real-time control of the bus system, we consider an economic objective function which includes the charging costs, a component related to service regularity, and an end-of-horizon term to account for the finite nature of the control horizon considered. Service irregularity can be seen as a cost for the transit agency since it can result in direct (e.g. fines on deviations from schedule) or indirect penalties (e.g. a decrease in bus ridership). 

With the modeling choices presented so far, the electric bus control problem can be expressed as the following optimization problem:
\begin{subequations} \label{eq:intractable_problem}
\begin{align}
    \text{min} \quad & J = p_{\text{el}} P_{\text{char}} \sum_{i,k,l} c_{l,i}^k \nonumber \\ 
    & \hspace{0.5cm} + p_{\text{wait}} \sum_{i,j,k,l} |t_{l,i,j}^k-t_{l,i-1,j}^k-H_l| \nonumber \\
    & \hspace{0.5cm} + p_{\text{end}}Q\sum_i \text{max}(0,\Bar{\sigma}-\sigma_{i,\text{end}}) \tag{\ref{eq:intractable_problem}} \\
    \text{s.t.} \quad & \: \eqref{eq:time_bounds}\textendash\eqref{eq:time_last},\eqref{eq:exclusion_1},\eqref{eq:exclusion_2},\nonumber
\end{align}
\end{subequations}
where the superscript $k$ denotes the visit index in case of repeated visits to the same stops by the same bus. The first term in the objective function is the total charging cost when electricity is purchased at the fixed price $p_{\text{el}}$. Note that this coefficient could in principle be adjusted to model time-of-use electricity pricing, e.g. by taking a different value for each visit index $k$ depending on the time of day. The second term penalizes deviations from the desired headway $H_l$ for each line $l$. These irregularities in bus service are given a monetary value with the cost term $p_{\text{wait}}$ to reflect their negative impact on passengers. 

Finally, the last term in the objective function of \eqref{eq:intractable_problem} is meant to mitigate end-of-horizon effects by applying a penalty for each bus $i$ whose state-of-charge at the end of the horizon $\sigma_{i,\text{end}}$ is lower than a desired value $\Bar{\sigma}$. The differences from this desired state-of-charge are expressed as energy quantities and given a price $p_{\text{end}}$. This last objective term is necessary to ensure that buses do not empty their batteries too early on in the day, which can happen if the horizon is short enough that buses can reach its end without needing to use the chargers. The design of $\Bar{\sigma}$ and $p_{\text{end}}$ is more open than for the other terms and should reflect the trade-off between the immediate costs incurred by charging at the present time and the uncertainty associated with a delayed charging. This aspect is detailed further in the case study below.

For numerical reasons, it is desirable to reformulate \eqref{eq:intractable_problem} as an equivalent MILP by expressing the 1-norm and max function used in the objective function differently. This is done by introducing continuous slack variables for some of the objective terms. This results in the following MILP:
\begin{subequations} \label{eq:MILP}
\begin{align}
    \text{min} \quad & J = p_{\text{el}} P_{\text{char}} \sum_{i,k,l} c_{l,i}^k + p_{\text{wait}} \sum_{i,j,k,l} \eta_{l,i,j}^k \nonumber \\ 
    & \hspace{0.5cm} + p_{\text{end}}Q\sum_i \nu_i \tag{\ref{eq:MILP}} \\
    \text{s.t.} \quad & \: \eqref{eq:time_bounds}\textendash\eqref{eq:time_last},\eqref{eq:exclusion_1},\eqref{eq:exclusion_2}, \nonumber \\
    & \eta_{l,i,j}^k \geq t_{l,i,j}^k-t_{l,i-1,j}^k-H_l, \nonumber \\
    & \eta_{l,i,j}^k \geq H_l-t_{l,i,j}^k+t_{l,i-1,j}^k, \nonumber \\
    & \nu_i \geq 0, \nonumber \\
    & \nu_i \geq \Bar{\sigma}-\sigma_{i,\text{end}}, \nonumber
\end{align}
\end{subequations}
where $\eta$ and $\nu$ denote the slack variables in question.

Mixed-integer programs like \eqref{eq:MILP} can be solved to optimality by dedicated solvers using the branch-and-bound method, such as Gurobi. The electric bus controller operates in a receding horizon fashion by assembling and solving \eqref{eq:MILP} at regular intervals. The optimal control inputs obtained from solving one instance of the MILP are applied until the next instance is solved. This process is repeated until the end of the total control period (e.g. one day of bus operations).

\begin{rem}
The proposed control method is built on a rather abstract representation of a bus network in order to allow for long time horizons. It should therefore be seen as a high-level control layer able to provide travel time, holding, and charging command references to individual buses. In practical situations, this high-level layer should be complemented with a low-level layer to control vehicle actuation, or to provide guidance to the driver. The design of such controllers is left outside of the scope of this paper, however.
\end{rem}

\section{Simulations}

The proposed controller is now tested in a simulation framework aiming to reproduce realistic bus operations. The results presented hereafter focus mainly on conflict avoidance at the chargers.

\subsection{Settings}

Real-life data from a bus line in Gothenburg, Sweden, is used to construct a small network with two bus lines. More information on the dataset from which this network is adapted can be found in \cite{lacombe}. The central station of the city is chosen as the common terminal, and is assumed to host a single fast charger shared by these two lines.

Based on real-life timetables, the desired headway for both lines is set at 5 minutes. The numbers of bus stops of these two lines are ${m_1=24}$ and ${m_2=30}$, and the numbers of buses are chosen as ${n_1=5}$ and ${n_2=6}$. The total simulation time, noted $T_{\text{sim}}$, is meant to represent one full day of bus operations and is set to 14 hours (e.g. from 6 am to 8 pm). The proposed controller computes updated control inputs every 5 minutes. The minimum state-of-charge allowed for both lines is set to ${\sigma^{\text{min}}=0.3}$. Additional modeling parameters are given in Table \ref{tb:parameters}. 

\begin{table}[hb]
\begin{center}
\caption{Simulation parameters.}
\label{tb:parameters}
\begin{tabular}{llllll} \hline
$Q$ & \hspace{-0.4cm} $=264$ kWh & $\sigma^{\text{min}}$ & \hspace{-0.4cm} $=0.3$ & $p_{\text{el}}$ & \hspace{-0.4cm} $=0.08$ \euro/kWh \\ 
$P_{\text{char}}$ & \hspace{-0.4cm} $=300$ kW & $T$ & \hspace{-0.4cm} $=60$ min & $p_{\text{wait}}$ & \hspace{-0.4cm} $=0.0025$ \euro/s \\ 
$d_{\text{pax}}$ & \hspace{-0.4cm} $=1.5$ s &  $T_{\text{sim}}$ & \hspace{-0.4cm} $=840$ min & $p_{\text{end}}$ & \hspace{-0.4cm} $=5\,p_{\text{el}}$ \\ 
$d_{\text{char}}$ & \hspace{-0.4cm} $=10$ s & $M$ & \hspace{-0.4cm} $=10^5$ & & \\
\hline
\end{tabular}
\end{center}
\end{table}

Unlike the bus network model presented earlier, which is completely deterministic, the simulation framework contains stochastic elements in order to mimic real public transit operations. The accumulation of passengers at each stop $j$ of line $l$ is sampled from a Poisson process using $\lambda_{l,j}$ as a parameter. Similarly, the influence of traffic on the $j$-th link of line $l$ is modeled by taking samples from a log-normal distribution centered on $T^{\text{min}}_{l,j}$ (\cite{hans}). These samples limit the maximum speed at which buses can travel on each link and are generated every time a bus is leaving a stop.

Concerning the terminal cost in \eqref{eq:intractable_problem}, we propose here a simple design rule for $\Bar{\sigma}$. It seems desirable for buses to end their service with a state-of-charge close to the minimum allowed in order to benefit from the cheaper overnight charging alternative. One option is then to have the state-of-charge target decrease linearly during the day, from a full battery charge initially to $\sigma^{\text{min}}$ at the end of the day. In other words, at any simulation time $t$, we have:  
\begin{equation}
    \Bar{\sigma}(t) = \sigma^{\text{min}} + \frac{T_{\text{sim}} - T - t}{T_{\text{sim}}}(1-\sigma^{\text{min}}).
\end{equation}
As for the cost of deviations from $\Bar{\sigma}$, it can be chosen as several times $p_{\text{el}}$ in order to incentivize sufficient charging. Here, we chose $p_{\text{end}}=5\,p_{\text{el}}$. 

As a way to evaluate the proposed controller, let us now consider a simple first-come, first-served charging heuristic. With this baseline controller, every bus starts charging immediately upon reaching the terminal, or as soon as the charger becomes available in case another bus is already charging. Note that this simple scheme might not be too different from what most transit agencies are using today. Here, each bus charges until the current state-of-charge target $\Bar{\sigma}$ is reached. In order to guarantee fair comparisons, the baseline controller is allowed to take \textit{locally optimal} control actions in terms of service regularity. That is, travel time commands are chosen so as to be closest to headway satisfaction at the next stop, and terminal holding is such that buses are dispatched according to the desired headway whenever possible. However, note that these control actions are purely \textit{reactive}, unlike those taken by the proposed controller. 

Both control methods are run with the same random seed in each simulation. In order for bus operations to stabilize from their initial conditions, costs only start to be counted after the first 30 minutes of simulation. Every bus is initialized with a fully charged battery. 

\subsection{Results}

Table \ref{tb:results} presents the main simulation results for a day of bus operations, where the values given are averages obtained from 25 independent repetitions. In these simulations, each version of MILP \eqref{eq:MILP} tends to have around 2500 decision variables, and the average time to solve each problem to optimality is $5$ s. 

\begin{table}[hb]
\begin{center}
\caption{Results.}
\label{tb:results}
\begin{tabular}{|c|c|c|c|} \hline
& Service Cost & Charging cost & Total cost \\ \hline 
Controller & 1102 \euro & 239 \euro & 1341 \euro \\ 
Baseline & 1306 \euro & 257 \euro & 1563 \euro \\ \hline 
\end{tabular}
\end{center}
\end{table}

\begin{figure}[!t]
\centering
\subfloat[Proposed controller.]{\includegraphics[width=8.4cm]{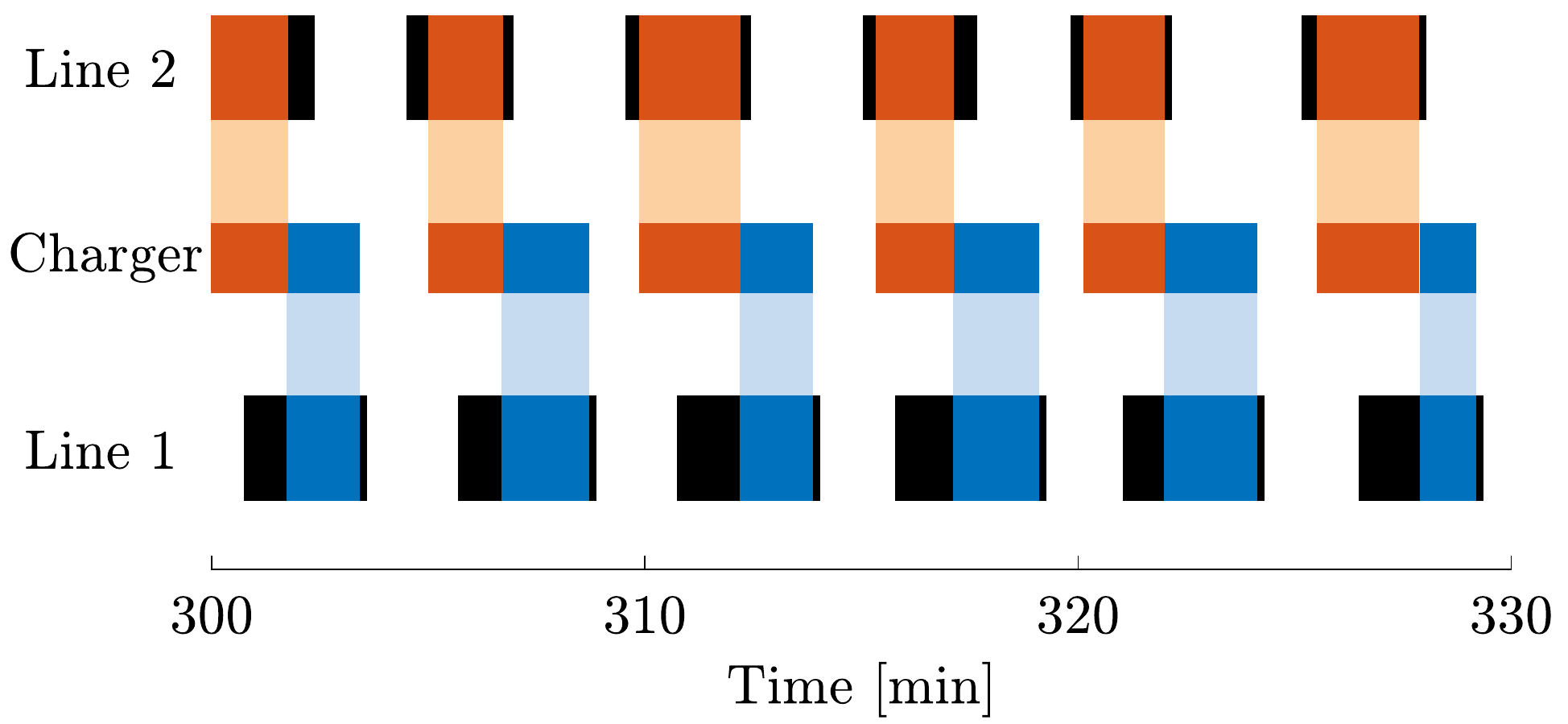}\label{fig:MILP}}%
\vfil
\subfloat[Baseline controller]{\includegraphics[width=8.4cm]{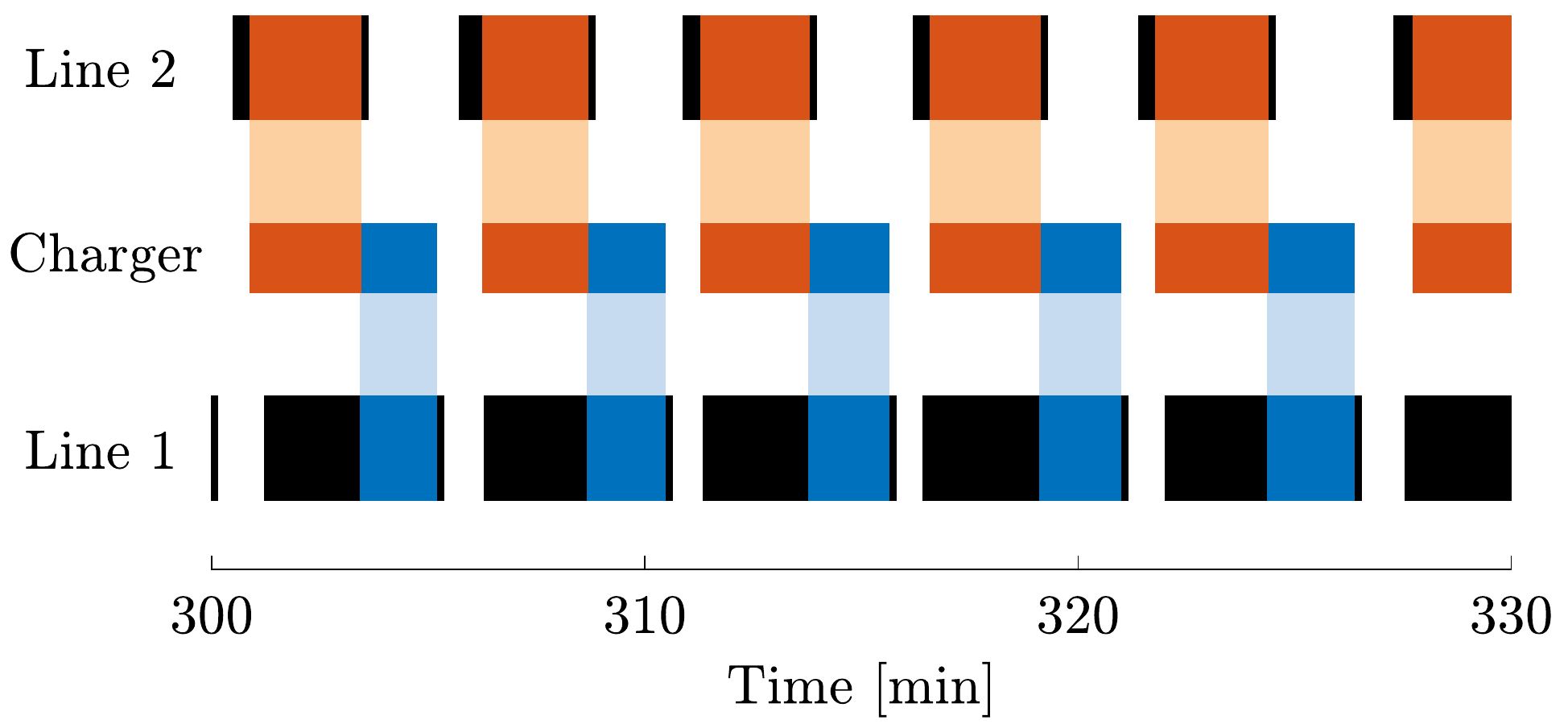}\label{fig:baseline}}%
\caption{Example of charger use during 30 minutes of simulation time. Each row represents the timeline of one bus line, or the charger timeline for the central row. In each bus line row, color rectangles indicate charger use, while black rectangles indicate that a bus from that line is at the terminal but not charging.} 
\label{fig:conflicts}
\end{figure}

It can be observed that the proposed controller achieves a total cost reduction of about $14\%$ compared with the baseline controller. Most of it comes from the service regularity cost, which is about $16\%$ lower for the proposed controller. This was found to be mainly due to the look-ahead feature of the proposed controller, and more precisely to its ability to anticipate and avoid charging conflicts at the terminal. The first-come, first-served baseline, on the other hand, often ends up in situations where two buses arrive simultaneously at the terminal. This usually causes one of them to wait for longer than desired and be dispatched late from the terminal, thus resulting in increased service irregularities at stops downstream.

Fig. \ref{fig:conflicts}. offers insights on the difference in the charging behavior of the two controllers. More often than not, only one bus is present at a time at the terminal with the proposed controller, whereas it is usually the case that two buses or more are at the terminal at the same time with the baseline controller. Baseline buses often block one another from using the charger as a result, which leads to long waiting periods materialized by the wide black rectangles in Fig. \ref{fig:baseline}. By contrast, the black rectangles are much narrower for buses in Fig. \ref{fig:MILP}., which manage to adapt their speed around the terminal in order to arrive when the charger becomes available. More generally, each baseline bus spends on average 78 s waiting at the terminal without charging during each visit. This figure drops to only 48 s per visit and per bus with the proposed controller.

A perhaps unexpected consequence of frequent charging conflicts is also an increase in energy consumption. Indeed, baseline buses spend more time waiting at the terminal due to charging conflicts: 18 hours of cumulative dwell time at the terminal on average, compared with only 15 hours with the proposed controller. This leads baseline buses to have a higher average speed when driving in order to travel with the desired headways, which results in increased charging needs, and hence in a higher charging cost for the baseline controller. This explains why the charging cost of the proposed controller is about $7\%$ lower than that of the baseline controller in Table \ref{tb:results}. 

\section{Conclusion}
We have presented in this work an approach to model and solve the integrated charging scheduling and real-time control problem for a network of electric buses. Under some general modeling assumptions, the control problem was formulated as an MILP. It was found in simulations that conflicts when using shared chargers can lead to increased operational costs by degrading service regularity and schedule satisfaction. Moreover, charging conflicts were also found to increase energy consumption as delayed buses need to drive faster in order to catch up with their schedule. In this context, it seems meaningful to use look-ahead optimization-based methods, like the one proposed in this paper, to reduce the occurrence of charging conflicts. A priority for future work would now be to focus on relaxing some of the modeling assumptions made in this paper. This would lead both to more accurate bus line models, by e.g. modeling bus capacity constraints, and to a more realistic representation of electric bus operations, by e.g. considering time-of-use electricity pricing. 

\bibliography{root} 

\begin{thebibliography}{11}
\providecommand{\natexlab}[1]{#1}
\providecommand{\url}[1]{\texttt{#1}}
\providecommand{\urlprefix}{URL }
\expandafter\ifx\csname urlstyle\endcsname\relax
  \providecommand{\doi}[1]{doi:\discretionary{}{}{}#1}\else
  \providecommand{\doi}{doi:\discretionary{}{}{}\begingroup
  \urlstyle{rm}\Url}\fi

\bibitem[{{Bloomberg New Energy Finance}(2018)}]{bloomberg}
{Bloomberg New Energy Finance} (2018).
\newblock Electric buses in cities: Driving towards cleaner air and lower
  {CO}2.
\newblock \emph{Bloomberg Finance LP}.

\bibitem[{Gkiotsalitis and Cats(2021)}]{gkiotsalitis3}
Gkiotsalitis, K. and Cats, O. (2021).
\newblock At-stop control measures in public transport: Literature review and
  research agenda.
\newblock \emph{Transportation Research Part E: Logistics and Transportation
  Review}, 145, 102176.

\bibitem[{Hans et~al.(2015)Hans, Chiabaut, and Leclercq}]{hans}
Hans, E., Chiabaut, N., and Leclercq, L. (2015).
\newblock Investigating the irregularity of bus routes: highlighting how
  underlying assumptions of bus models impact the regularity results.
\newblock \emph{Journal of Advanced Transportation}, 49(3), 358--370.

\bibitem[{He et~al.(2020)He, Liu, and Song}]{he}
He, Y., Liu, Z., and Song, Z. (2020).
\newblock Optimal charging scheduling and management for a fast-charging
  battery electric bus system.
\newblock \emph{Transportation Research Part E: Logistics and Transportation
  Review}, 142, 102056.

\bibitem[{Ibarra-Rojas et~al.(2015)Ibarra-Rojas, Delgado, Giesen, and
  Mu{\~n}oz}]{ibarra}
Ibarra-Rojas, O.J., Delgado, F., Giesen, R., and Mu{\~n}oz, J.C. (2015).
\newblock Planning, operation, and control of bus transport systems: A
  literature review.
\newblock \emph{Transportation Research Part B: Methodological}, 77, 38--75.

\bibitem[{Lacombe et~al.(2022)Lacombe, Gros, Murgovski, and
  Kulcs{\'a}r}]{lacombe}
Lacombe, R., Gros, S., Murgovski, N., and Kulcs{\'a}r, B. (2022).
\newblock Bilevel optimization for bunching mitigation and eco-driving of
  electric bus lines.
\newblock \emph{IEEE Transactions on Intelligent Transportation Systems},
  23(8), 10662--10679.

\bibitem[{Lajunen and Lipman(2016)}]{lajunen}
Lajunen, A. and Lipman, T. (2016).
\newblock Lifecycle cost assessment and carbon dioxide emissions of diesel,
  natural gas, hybrid electric, fuel cell hybrid and electric transit buses.
\newblock \emph{Energy}, 106, 329--342.

\bibitem[{Pelletier et~al.(2019)Pelletier, Jabali, Mendoza, and
  Laporte}]{pelletier}
Pelletier, S., Jabali, O., Mendoza, J.E., and Laporte, G. (2019).
\newblock The electric bus fleet transition problem.
\newblock \emph{Transportation Research Part C: Emerging Technologies}, 109,
  174--193.

\bibitem[{Perumal et~al.(2022)Perumal, Lusby, and Larsen}]{perumal}
Perumal, S.S., Lusby, R.M., and Larsen, J. (2022).
\newblock Electric bus planning \& scheduling: A review of related problems and
  methodologies.
\newblock \emph{European Journal of Operational Research}, 301(2), 395--413.

\bibitem[{Tang et~al.(2019)Tang, Lin, and He}]{tang}
Tang, X., Lin, X., and He, F. (2019).
\newblock Robust scheduling strategies of electric buses under stochastic
  traffic conditions.
\newblock \emph{Transportation Research Part C: Emerging Technologies}, 105,
  163--182.

\bibitem[{Xylia et~al.(2017)Xylia, Leduc, Patrizio, Kraxner, and
  Silveira}]{xylia}
Xylia, M., Leduc, S., Patrizio, P., Kraxner, F., and Silveira, S. (2017).
\newblock Locating charging infrastructure for electric buses in stockholm.
\newblock \emph{Transportation Research Part C: Emerging Technologies}, 78,
  183--200.

\end{thebibliography}


\end{document}